\documentclass[]{aa}
\usepackage{graphicx}
\begin{document}

\title{CCD photometric search for peculiar stars in open clusters. V. 
NGC~2099, NGC~3114, NGC~6204, NGC~6705 and NGC~6756\thanks{Based on 
observations obtained at Complejo Astron\'omico el Leoncito (CASLEO), 
operated under the agreement between the
Consejo Nacional de Investigaciones Cient\'ificas y T\'ecnicas de la Rep\'ublica 
Argentina and the National Universities of La Plata, C\'ordoba y San Juan;
ESO-La Silla, UTSO-Las Campanas and L.~Figl Observatory, Mt. Sch\"opfl (Austria)}}
\author{E.~Paunzen\inst{1,2}, O.I.~Pintado\inst{3,}\thanks
{Member of Carrera del Investigador del Consejo Nacional de 
Investigaciones Cient\'{i}ficas y T\'{e}cnicas de la Rep\'{u}blica 
Argentina and Visiting Astronomer at Complejo Astron\'omico El Leoncito 
operated under agreement between Consejo Nacional de Investigaciones 
Cient\'{i}ficas y T\'{e}cnicas de la Rep\'ublica Argentina and the 
National Universities of La Plata, C\'ordoba y San Juan}, H.M.~Maitzen\inst{1}}

\mail{Ernst.Paunzen@univie.ac.at}

\institute{Institut f\"ur Astronomie der Universit\"at Wien,
           T\"urkenschanzstr. 17, A-1180 Wien, Austria
\and       Zentraler Informatikdienst der Universit\"at Wien,
           Universit\"atsstr. 7, A-1010 Wien, Austria
\and	   Departamento de F\'isica, Facultad de Ciencias Exactas 
           y Tecnolog\'ia, Universidad Nacional de Tucum\'an, Argentina - Consejo Nacional 
		   de Investigaciones Cient\'ificas y T\'ecnicas de la Rep\'ublica Argentina}

\date{Received 2003; accepted 2003}
\authorrunning{E. Paunzen et al.}{}
\titlerunning{Photometric search for peculiar stars in open clusters. V.}{}

\abstract{We have investigated 1008 objects in the area of five
intermediate age
open clusters (NGC~2099, NGC~3114, NGC~6204, NGC~6705 and NGC~6756) 
via the narrow band $\Delta a$-system. The
detection limit for photometric peculiarity is very low (always less than
0.009\,mag) due to the high number of individual frames used (193 in total).
We have detected six peculiar objects in NGC 6705 and NGC 6756 from which
one in the latter is almost certainly an unreddened late type foreground star.
The remaining five stars are probably cluster members and
bona fide chemically peculiar objects (two are $\lambda$ Bootis
type candidates).
Furthermore, we have investigated NGC 3114, a cluster
for which already photoelectric $\Delta a$-measurements exist. A comparison
of the CCD and photoelectric values shows very good agreement. Again,
the high capability of our CCD $\Delta a$-photometric system to sort out
true peculiar objects together with additional
measurements from broad or intermediate band photometry is demonstrated.
}
\maketitle

\section{Introduction}

In continuation of our previous four papers dedicated to the search
for chemically peculiar objects via CCD $\Delta a$-photometry (Paunzen et al. 2002b),
another five open clusters in our Milky Way have been investigated. 

We present $\Delta a$-photometry for NGC~2099, NGC~3114, NGC~6204, 
NGC~6705 and NGC~6756 resulting in the detection of six peculiar objects from
which one turned out to be a late type foreground object.

The data of NGC 3114 allowed us to compare photoelectric measurements to those
of the CCD system showing excellent agreement.

We discuss the individual clusters and their peculiar objects including Johnson
$UBV$ and Str\"omgren $uvby$ photometry.

\begin{figure*}
\begin{center}
\includegraphics[width=155mm]{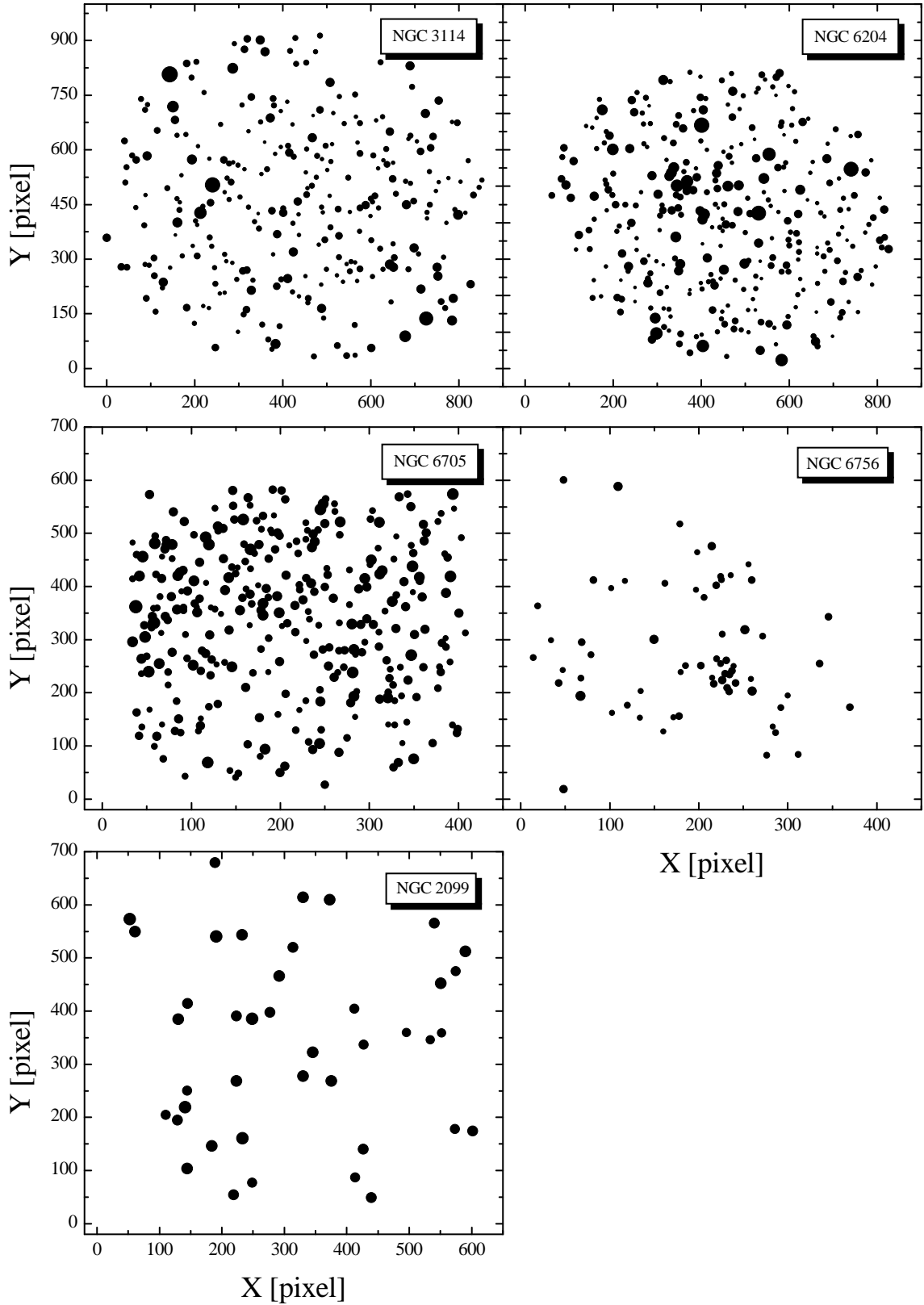}
\caption{Finding charts for the program clusters.
North is to the right and west
is upwards; 1 pixel\,=\,0.5$\arcsec$. The sizes (by area) of the open circles
are inversely proportional to the $V$-magnitudes taken from Tables 5 to 9
larger open circles denote brighter objects. We have plotted the clusters
sorted for the individual field of views.}
\label{charts}
\end{center}
\end{figure*}

\begin{figure*}
\begin{center}
\includegraphics[width=160mm]{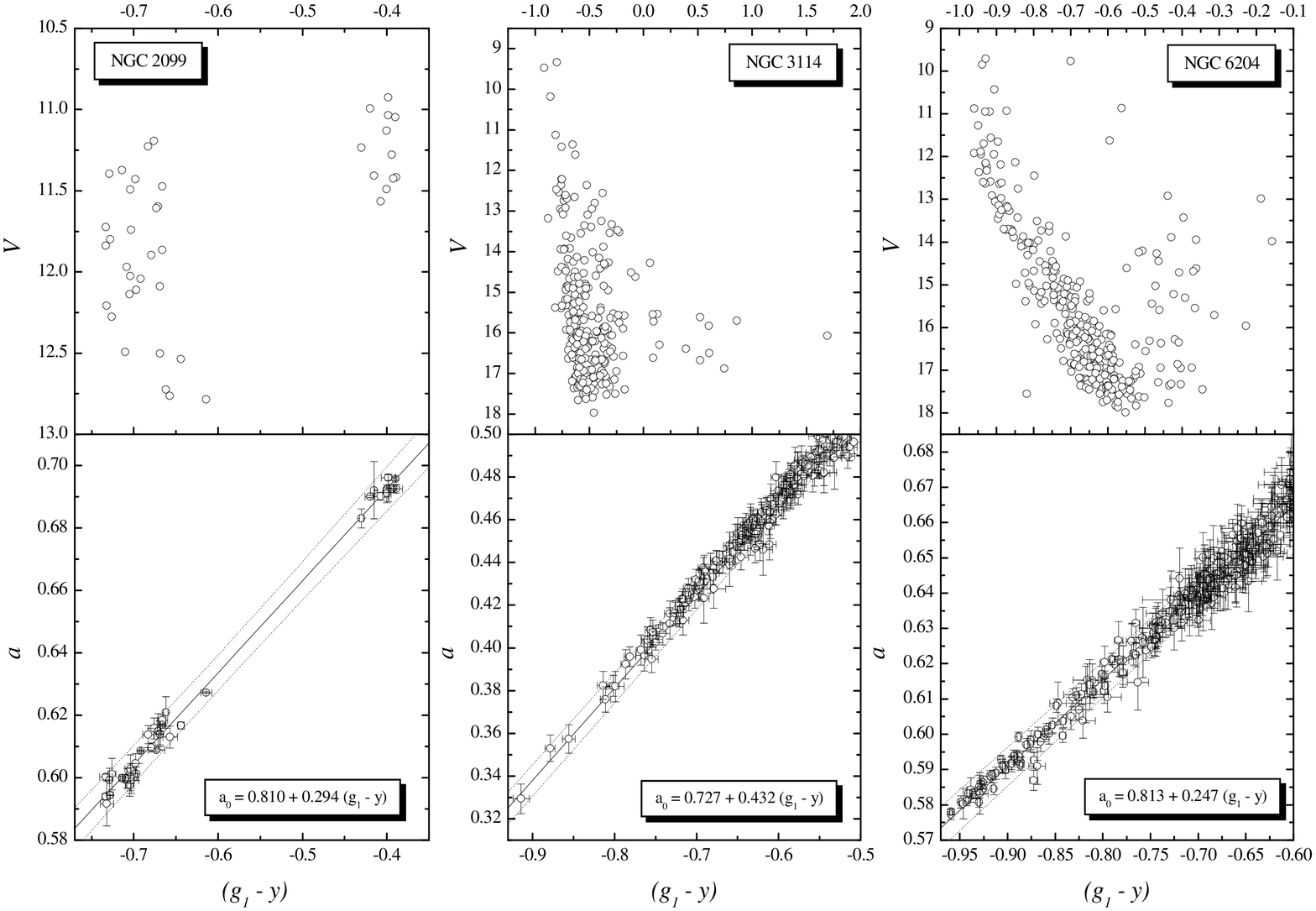}
\includegraphics[width=160mm]{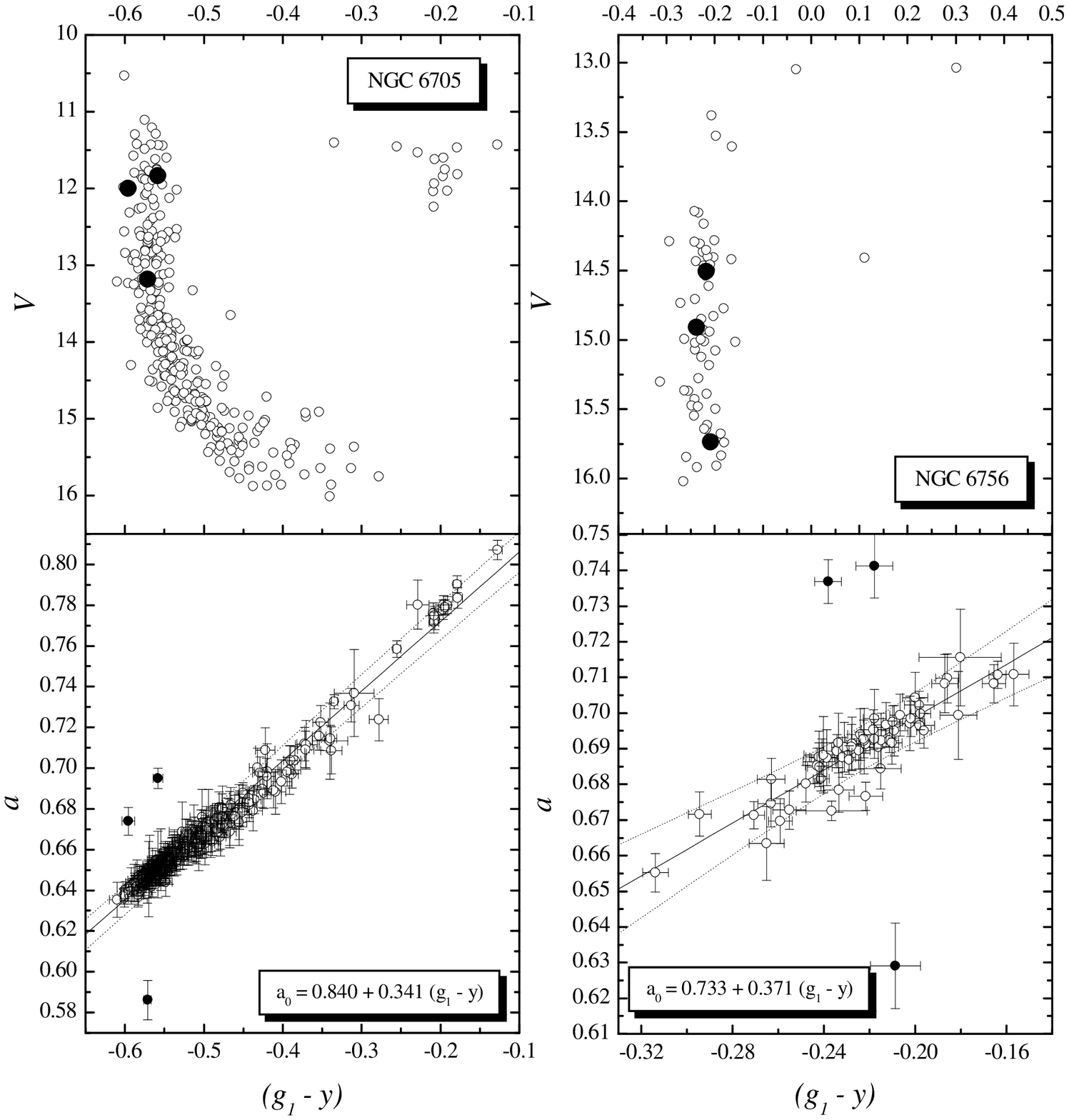}
\caption[]{Observed diagrams for our program clusters.
The solid line is the
normality line whereas the dotted lines are the confidence intervals
corresponding to 99.9\,\%. The error bars for each individual object
are the mean errors. The measurement errors of $V$ are much smaller than the
symbols and have been omitted. The scales for the upper and lower diagrams of NGC
3114, NGC 6204 and NGC 6756 are different because the relevant
range for peculiar objects in the $a$ versus $(g1-y)$ diagrams seemed
worthwhile to be shown.}
\label{all_plot}
\end{center}
\end{figure*}

\section{Observations, reduction and results}

Observations of the five open clusters were performed 
with the Bochum 61\,cm (ESO-La Silla), the Helen-Sawyer-Hogg 
61\,cm telescope (UTSO-Las Campanas Observatory),  
the 2.15\,m telescope at the Complejo Astron\'omico el Leoncito (CASLEO)
and the L. Figl Observatory (FOA) with the 150\,cm telescope
on Mt. Sch\"opfl (Austria) using the multimode instrument
OEFOSC. The characteristic of the instruments and field-of-views
can be found in Bayer et al. (2000) and Paunzen et al. (2002b).
The observing log is listed in Table \ref{log}.

The basic CCD reductions and a point-spread-function-fitting 
were carried out within standard IRAF routines.
The way of calculating the normality line, deriving the errors
and calibration of our $y$ measurements, is the same as in
previous works (Paunzen et al. 2002b). 

\begin{table}[t]
\begin{center}
\caption{Observing log}
\label{log}
\begin{tabular}{lcccccc}
\hline
Cluster & Site & Date & \#$_{N}$ & \#$_{g_{\rm 1}}$ & \#$_{g_{\rm 2}}$ & \#$_{y}$ \\
\hline
NGC 2099 & FOA & 03.1998 & 1 & 2 & 2 & 2 \\
NGC 3114 & CASLEO & 11.1998 & 4 & 14 & 17 & 19 \\
NGC 6204 & CASLEO & 08.2001 & 2 & 20 & 15 & 20 \\
NGC 6705 & ESO & 05.1995 & 5 & 15 & 15 & 13 \\
NGC 6756 & ESO & 05.1995 & 4 & 11 & 13 & 11 \\
& UTSO & 05.1995 & 1 & 1 & 2 & 1 \\
\hline
\end{tabular}
\end{center}
\end{table}

Table \ref{all_res} lists our observed open clusters and
their characteristics from the literature (as listed in the corresponding 
sections). Furthermore, the number of observed stars, the number of individual
frames (Table \ref{log}), the regression coefficients for all
transformations and normality lines  
as well as the 3$\sigma$ detection limits are given.

The finding charts (only available in the electronic form)
of our open clusters are shown in Fig. \ref{charts}.
We have grouped the charts according to the
telescopes used and the very different field of views.

\begin{table*}[t]
\begin{center}
\caption{Summary of results and parameters from the literature. 
The errors in the final digits of the corresponding quantity
are given in parenthesis.}
\label{all_res}
\begin{tabular}{lccccc}
\hline
Name & NGC 2099 & NGC 3114 & NGC 6204 & NGC 6705 & NGC 6756 \\
     & C0549+325 & C1001$-$598 & C1642$-$469 & C1848$-$063 & C1906+046 \\
\hline
$l/b$ & 178/+3 & 283/$-$4 & 339/$-$1 & 27/$-$3 & 39/$-$2 \\
$E(B-V)$ & 0.30 & 0.07 & 0.45 & 0.43 & 0.91 \\
$d$\,[pc] & 1400 & 920 & 1200 & 1900 & 3300 \\
log\,$t$ & 8.60 & 8.48 & 8.30 & 8.40 & 8.11 \\
Tr-type & II 1 r & II 3 r & I 3 m & I 2 r & I 2 m \\
n(obj) & 41 & 271 & 319 & 312 & 65 \\
$\Delta$a/$(B-V)_{0}$ & & & & +0.046 $-$0.047 \\
& & & & +0.026 $-$0.046 \\
& & & & $-$0.063 $-$0.050                                                         \\
$\Delta$a/$(b-y)_{0}$ & & & & & +0.051 $-$0.069 \\
& & & & & +0.053 $-$0.060 \\
& & & & & $-$0.067 $-$0.024 \\
$V$\,=\,a\,+\,b$\cdot(y)$ & $-$1.51(23) 1.01(2) & $-$4.69(7) 1.008(3) & 
$-$4.75(8) 0.987(4) & $-$8.38(5) 1.009(2) & $-$7.42(15) 0.962(7) \\
$a_{0}$\,=\,a\,+\,b$\cdot(g_{1}-y)$ & 0.810(2) 0.294(4) & 0.727(1) 0.432(1) &
0.813(1) 0.247(2) & 0.840(3) 0.341(7) & 0.773(5) 0.371(22) \\
$a_{0}$\,=\,a\,+\,b$\cdot(B-V)$ & 0.575(2) 0.097(3) & 0.353(6) 0.164(6) & 0.545(2) 0.103(2)&
0.602(1) 0.122(2) & \\
$a_{0}$\,=\,a\,+\,b$\cdot(b-y)$ & & & & & 0.546(17) 0.212(25) \\
3\,$\sigma$\,[mag] & 0.009 & 0.008 & 0.007 & 0.008 & 0.009 \\
n(frames) & 6 & 50 & 55 & 43 & 39 \\
\hline
\end{tabular}
\end{center}
\end{table*}

\begin{table*}[t]
\caption{Stars of the investigated open clusters with peculiar $\Delta a$-values;
No. 17 of NGC 6756 is most probably an unreddened foreground star.}
\label{results}
\begin{center}
{\tiny
\begin{tabular}{crrrrcccccccccc}
\hline
Cluster & 
\multicolumn{1}{c}{$No_{1}$} & \multicolumn{1}{c}{$No_{2}$} & 
\multicolumn{1}{c}{$X$} & \multicolumn{1}{c}{$Y$} &
\multicolumn{1}{c}{$V$}&\multicolumn{1}{c}{$a$}&\multicolumn{1}{c}{$\sigma$}&
\multicolumn{1}{c}{$(g_{1}-y)$}&\multicolumn{1}{c}{$\sigma$}&
\multicolumn{1}{c}{$\Delta a$}&
\multicolumn{1}{c}{$(B-V)$}&\multicolumn{1}{c}{$\Delta a$} &
\multicolumn{1}{c}{$(b-y)$}&\multicolumn{1}{c}{$\Delta a$} \\   
\hline
NGC 6705
& 51	& 7730	& 85.1	& 419.6	& 11.84 &	0.695	& 0.005	& $-$0.558 &	0.006	& +0.045	&
+0.383	& +0.046	\\
& 56	& 7777	& 91.2	& 429.9	& 13.18 &	0.586	& 0.010	& $-$0.571 &	0.005	& $-$0.059	&
+0.384	& $-$0.063	\\
& 215	& 8157	& 267.5	& 520.4	& 12.00 &	0.674	& 0.007	& $-$0.596 &	0.008	& +0.037	&
+0.380	& +0.026	\\
NGC 6756
& (17)	& 33	& 120.0	& 176.3	& 14.91 &	0.737	& 0.006	& $-$0.238 &	0.006	& +0.052	&
& & +0.661	& +0.051	\\
& 46	& 24	& 234.9	& 233.4	& 14.51 &	0.741	& 0.009	& $-$0.218 &	0.008	& +0.049	&
& & +0.670	& +0.053	\\
& 52	& 66	& 256.5	& 440.6	& 15.74 &	0.629	& 0.012	& $-$0.209 &	0.011	& $-$0.067	&
& & +0.706	& $-$0.067	\\
\hline
\\
\multicolumn{13}{l}{Col. 1: Cluster name}\\
\multicolumn{13}{l}{Col. 2: Notation sorted after $X$ and $Y$, respectively (Fig.
\ref{charts})}\\
\multicolumn{13}{l}{Col. 3: Notation according to Sung et al. (1999; NGC 6705),
Delgado et al. (1997; NGC 6756)} \\
\multicolumn{13}{l}{Col. 4, 5: $X$ and $Y$ coordinates in the finding charts (Fig.
\ref{charts})} \\
\multicolumn{13}{l}{Col. 6: Johnson $V$ magnitude} \\
\multicolumn{13}{l}{Col. 7, 8: mean $a$-index and its standard deviation} \\
\multicolumn{13}{l}{Col. 9, 10: mean $(g_{1}-y)$ value and its standard deviation} \\
\multicolumn{13}{l}{Col. 11: Deviation from cluster line 
 $a_{0}$\,=\,a\,+\,b$\cdot(g_{1}-y)$ using the
 corresponding constants as listed in Table \ref{all_res}} \\
\multicolumn{13}{l}{Col. 12: $(B-V)$ from the literature} \\
\multicolumn{13}{l}{Col. 13: Deviation from cluster line 
$a_{0}$\,=\,a\,+\,b$\cdot(B-V)$ using the
corresponding constants as listed in Table \ref{all_res}} \\
\multicolumn{13}{l}{Col. 14: $(b-y)$ from Delgado et al. (1997)} \\
\multicolumn{13}{l}{Col. 15: Deviation from cluster line 
$a_{0}$\,=\,a\,+\,b$\cdot(b-y)$ using the
corresponding constants as listed in Table \ref{all_res}} \\
\end{tabular}
}
\end{center}
\end{table*}

Tables 5 to 9 with all data for the individual cluster stars are available
in electronic form at the CDS via anonymous ftp to cdsarc.u-strasbg.fr (130.79.125.5),
http://cdsweb.u-strasbg.fr/Abstract.html
or upon request from the first author. These Tables include the cross
identification of objects from the literature, the observed $(g_{1}-y)$ and
$a$ values with their corresponding errors, $V$ magnitudes,
the $(B-V)$ or $(b-y)$ values from the literature,
$\Delta a$-values derived from the normality lines of $(g_{1}-y)$,
$(B-V)$ or $(b-y)$, and the number of observations, respectively.

The diagnostic diagrams for
all five open clusters are shown in Fig. \ref{all_plot}. Furthermore,
the normality lines and the confidence intervals corresponding to 99.9\,\%
are plotted.

All photometric data and limits are given in magnitudes.

\subsection{NGC 2099}

A comprehensive study of this open cluster is given by Nilakshi \& Sagar
(2002). They included all relevant sources and data from the literature.

We have observed only the brightest members ($V>13$) of the core region
and find no photometric CP candidate. Although we have only six frames in total,
the $\Delta a$-detection limit is very low (0.009) caused by the brightness
of the observed objects.

The Johnson $UBV$ measurements of the literature as listed in Table 1 of 
Nilakshi \& Sagar (2002) were averaged in order to derive mean photometric mean
values within this system.
The main sequence and the region of the red giants are clearly visible in Fig.
\ref{all_plot}.

\subsection{NGC 3114}

This cluster was already observed photoelectrically in the
$\Delta a$-system by Maitzen et al. (1988). They observed 127 stars of this
cluster with $V>12$ and found six CP2 stars. We have used NGC 3114 to further
test our new CCD photometric system and to expand the $V$-range of
observed members (it is also the less distant open clusters
investigated so far in the CCD $\Delta a$-system with 920\,pc). 
Unfortunately, all six bona-fide
CP stars are either outside our observed field of view or too bright for
CCD measurements. Nevertheless, we have seven stars in common with the
work of Maitzen et al. (1988) for which we can immediately check the
observed $\Delta a$-values. Admitting that this is
only a check for a null result in $\Delta a$ we notice a very good coincidence
of both sources/techniques.

The CCD Johnson $UBV$ for this cluster were taken from Sagar \& 
Sharples (1991) and Carraro \& Patat (2001). 

Figure \ref{all_plot} shows the $V$ versus $(g_1-y)$ diagram for NGC 3114.
The main sequence goes down to about 18th magnitude with many non-members
especially cooler than $(g_1-y)>-0.5$. We find no further photometric
CP stars for this cluster within a detection limit of 0.008. This
is probably caused by the temperature range of the observed members.
We find that $(g_1-y)=-0.69$ corresponds to $(B-V)_0=0.44$ or
a spectral type of about F5. Earlier than that, only 43 objects have been
observed (all other stars are cooler). Since the incidence of
CP stars significantly drops towards spectral types of A5,
the incidence of late type peculiar stars is rather improbable
(note that already six have been discovered before).

\subsection{NGC 6204}
 
NGC 6204 was often analyzed with its companion Hogg 22 which lies
only 6$\arcmin$ away (both are not associated with each other). 
The first investigation by Whiteoak (1963)
was substantiated by Moffat \& Vogt (1973). The latter give an earliest
spectral type of B8 and a distance of 810\,pc. Kjeldsen \& Frandsen (1991)
presented CCD Johnson $UBV$ photometry deriving a distance of 1200\,pc
and an age of 200\,Myr. Later, Forbes \& Short (1996) confirmed these results
using appropriate isochrones.  

Our observations cover the whole area of NGC 6204 as described in 
Kjeldsen \& Frandsen (1991) and Forbes \& Short (1996) spanning
a magnitude range of about nine magnitudes. Although
55 frames were observed in total resulting in a very low detection limit of
0.007, no photometric CP candidate was discovered.

\subsection{NGC 6705} 

The apparent radius of NGC 6705 (M11) is 16$\arcmin$ which is significantly
more than the field of view of the telescope (3$\arcmin$\,x\,4$\arcmin$) used
for this investigation. We have therefore observed only the innermost region.
Kjeldsen \& Frandsen (1991), 
Brocato et al. (1993) and Sung et al. (1999) presented extensive studies
based on CCD Johnson $UBVI$ photometry. Sung et al. (1999) found mass segregation
based on 166.491 stars brighter than 21th magnitude (see their
Figure 3). The slope of
the mass function increases systematically with increasing radius from the center.

We have checked the consistency of all published CCD photometric sources.
The photographic values listed by Johnson et al. (1956) and Mathieu (1984) were
not considered since such measurements are in generally not as accurate as CCD
ones. We find a general trend of the data listed by Brocato et al. (1993) with
respect to those of Kjeldsen \& Frandsen (1991) and Sung et al. (1999). The offset
is about 0.70 for Johnson $V$, furthermore a slope of 0.74 for Johnson
$(B-V)$ was found. The data of Brocato et al. (1993) were also checked against
the photographic ones by Johnson et al. (1956) and Mathieu (1984) yielding
essentially the same results. The correlations of all other data sources result
in offsets close to zero with a slope of one (within the error bars). For our
further analysis we have used photometric mean values from 
Kjeldsen \& Frandsen (1991) and Sung et al. (1999).

From Fig. \ref{all_plot} we find three photometric CP candidates
for NGC 6705. Two of them (No. 51 and 56; 7730 as well as 8157 according
to Sung et al. 1999) have significant positive $\Delta a$-values whereas
one (56; 7777) exhibit a negative one. Unfortunately, none of these objects
is included in the membership determination by McNamara et al. (1977).
All three stars have $(B-V)_0$ of $-$0.05 and $(U-B)_0$ values
from $-$0.07 to $-$0.17 resulting in a $Q$ value from $-$0.04 to
$-$0.14 (Golay 1974) qualifying them as bona fide
cluster members (Fig. \ref{all_plot}). These values are typically for spectral types of
B8 to A0 assuming a reddening of $E(B-V)=0.43$. 

According to the results
of Cowley et al. (1970) the photometric indices are in very good
agreement for late B type Si or Cr stars. If we assume $(V_0-M_V)=11.55$ (Sung et al.
1999) then absolute magnitudes of +0.3 and +0.5 for No. 51 and 215 can be
derived. Again, these absolute magnitudes are typical for CP stars in this
temperature range (Grenier et al. 1981, G\'omez et al. 1998).

Object No. 56 exhibits a significant negative $\Delta a$-value of
$-$0.059 which is typical for a Be or a very extreme $\lambda$ Bootis star.
An absolute magnitude of +1.6 would be still in the range of
classical field $\lambda$ Bootis stars (Paunzen et al. 2002a). However,
spectroscopic observations have to clarify the true nature of this object.

We also note that Johnson et al. (1956) classified one star (No. 7483;
Sung et al. 1999) as B3p (Blue Straggler?). Our photometry results
in a $\Delta a$-value of +0.003 which seems to rule out a peculiar nature.
But we have to emphasize that magnetic CP stars often show a variation of
the $\Delta a$-value over their magnetic cycle. On the other hand, the error
of the $a$-value itself (averaged over five different nights) is only 0.003
which might be another argument against chemically peculiarity.

\subsection{NGC 6756}

The only recent paper including NGC 6756 is that of Delgado et al. (1997).
They presented CCD Str\"omgren $uvby$ photometry of 368 stars in the
area of this open cluster. It is the most distant program cluster (3300\,pc)
which allowed us the cover the complete cluster area.

In total, three objects are photometrically peculiar: No. 17, 46 (33 and
24 according to Delgado et al. 1997, respectively; both
with positive $\Delta a$-values) and No. 52 (66; negative one).

For the following calculations we have used $E(b-y)=0.73$ and
$(V_0-M_V)=12.60$ as listed in Delgado et al. (1997).

The Str\"omgren $uvby$ indices for No. 46 further strengthen its chemically 
peculiar nature. 
We find $(b-y)_0$\,=\,$-$0.060, [$m_1$]\,=\,0.105, [$c_1$]\,=\,0.671 and
$M_{V}$\,=\,$-$0.42 which are typical for a B8\,Si star (Cameron 1966).

The values for the apparent peculiar object No. 17 are typical for
an unreddened late type star (Crawford \& Mander 1966) and rule out 
chemical peculiarity: $(b-y)$\,=\,0.661, [$m_1$]\,=\,0.502 and [$c_1$]\,=\,0.340. 
This case shows the importance of additional
photometric observations in a broad or intermediate band system.

Unfortunately, Delgado et al. (1997) have only measured $(b-y)$ for 
No. 52 making a final comment on its nature difficult. If it is a cluster
member, then $(b-y)_0$\,=\,$-$0.024 and $M_{V}$\,=\,+0.41
which is in the range of the $\lambda$ Bootis group. Again, further
photometric or spectroscopic data are needed to shed more light
on its nature.

\section{Conclusions}

We have presented high precision narrow band $\Delta a$-photometry for 
1008 objects from 193 individual frames for five intermediate aged open
clusters of the Milky Way. With a very low detection limit of less
equal than 0.009, we find six peculiar objects in NGC 6705 and NGC 6756.
One object in NGC 6756 turned out to be a late type foreground object.
NGC 3114 served as further test case to compare the new CCD with the
old photoelectric $\Delta a$-system. It has been shown that our narrow
band CCD $\Delta a$-photometric system together with additional
measurements of the Johnson $UBV$ ot Str\"omgren $uvby$ systems is highly 
efficient to detect peculiar objects on the upper main sequence.

\begin{acknowledgements}
The CCD and data acquisition system at CASLEO has been partly financed
by R.M. Rich through U.S. NSF Grant AST-90-15827.
This work benefitted from the financial contributions of the City of
Vienna
(Hochschuljubil\"aumsstiftung projects: Wiener Zweikanalphotometer and
H-112/95 Image Processing).
EP acknowledges partial support by the Fonds zur F\"orderung der
wissenschaftlichen Forschung, project P14984.
Use was made of the SIMBAD database, operated at CDS, Strasbourg, France and
the WEBDA database, operated at the Institute of Astronomy of the University
of Lausanne. This research has made use of NASA's Astrophysics Data System.
\end{acknowledgements}

\end{document}